\documentstyle[12pt]{article}

\newcommand{\om}{\omega}
\newcommand{\al}{\alpha}
\newcommand{\ep}{\epsilon}

\newcommand{\la}{\lambda}

\newcommand{\deebar}{\tilde{\partial}}
\newcommand{\df}{\stackrel{\rm def}{=}}

\newcommand{\lb}{\lbrack}
\newcommand{\rb}{\rbrack}

\newcommand{\msc}[1]{\mbox{\scriptsize #1}}
\newcommand{\dsp}{\displaystyle}

\newcommand{\bc}{\mbox{{\bf C}}}
\newcommand{\br}{\mbox{{\bf R}}}
\newcommand{\bz}{\mbox{{\bf Z}}}

\newcommand{\bv}{\mbox{{\bf v}}}
\newcommand{\bsv}{\msc{{\bf v}}}
\newcommand{\bsz}{\msc{{\bf Z}}}
\newcommand{\bsr}{\msc{{\bf R}}}

\newcommand{\cB}{{\cal B}}
\newcommand{\cC}{{\cal C}}

\newcommand{\cL}{{\cal L}}

\newcommand{\cD}{{\cal D}}
\newcommand{\cS}{{\cal S}}

\newcommand{\ket}[1]{{|#1\rangle}}
\newcommand{\bra}[1]{{\langle#1|}}

\newcommand{\Th}[2]{\Theta_{#1,#2}}
\newcommand{\th}{{\theta}}

\newcommand{\tr}{\mbox{Tr}}
\newcommand{\mod}{\mbox{mod}}

\newcommand{\lsim}{\stackrel{<}{\sim}}
\newcommand{\gsim}{\stackrel{>}{\sim}}

\newcommand{\ttau}{\tilde{\tau}}

\newcommand{\tz}{\tilde{z}}
\newcommand{\tU}{\tilde{U}}
\newcommand{\tw}{\tilde{w}}

\newcommand{\nn}{\nonumber\\}

\newcommand {\eqn}[1]{(\ref{#1})}

\makeatletter
\@addtoreset{equation}{section}

\makeatother

\begin{document}

\vskip 7mm
\begin{titlepage}
 
 \renewcommand{\thefootnote}{\fnsymbol{footnote}}
 \font\csc=cmcsc10 scaled\magstep1
 {\baselineskip=14pt
 \rightline{
 \vbox{\hbox{hep-th/0107189}
       \hbox{UT-952}
       }}}

 \vfill
 \baselineskip=20pt
 \begin{center}
 \centerline{\Huge  Boundary States of D-branes in $AdS_3$ } 
 \vskip 5mm 
 \centerline{\Huge  Based on Discrete Series}

 \vskip 2.0 truecm

\noindent{\it \large Yasuaki Hikida and Yuji Sugawara} \\
{\sf hikida@hep-th.phys.s.u-tokyo.ac.jp~,~
sugawara@hep-th.phys.s.u-tokyo.ac.jp}
\bigskip

 \vskip .6 truecm
 {\baselineskip=15pt
 {\it Department of Physics,  Faculty of Science, \\
  University of Tokyo \\
  Hongo 7-3-1, Bunkyo-ku, Tokyo 113-0033, Japan}
 }
 \vskip .4 truecm

 \end{center}

 \vfill
 \vskip 0.5 truecm

\begin{abstract}
\baselineskip 6.7mm

We study D-branes in the Lorentzian $AdS_3$ background 
from the viewpoint of boundary states, emphasizing the role 
of open-closed duality in string theory. 
Employing the world sheet with Lorentzian signature,
we construct the Cardy states with the discrete series.
We show that they are compatible with (1) unitarity and 
normalizability, and (2) the spectral flow symmetry, in the open 
string spectrum. We also discuss their brane interpretation.
We further show that in the case of superstrings on 
$AdS_3\times S^3 \times T^4$, our Cardy states yield  an infinite 
number of physical BPS states in the open string channel, on which 
the spectral flows act consistently.

\end{abstract}

\setcounter{footnote}{0}
\renewcommand{\thefootnote}{\arabic{footnote}}
\end{titlepage}

\newpage

\section{Introduction}

Closed string theory on the $AdS_3$ background has
attracted a great deal of attention for several reasons. 
Among other things, it provides a non-trivial example of 
solvable string theories on a non-compact curved space-time,
and it was studied for this reason in several pioneering works \cite{sl2}.
More recently, it has been studied from the viewpoint of  
the AdS/CFT correspondence \cite{AdSCFT} at the stringy level 
in Refs.~\cite{GKS,DORT,KS} and also in many 
subsequent works. A detailed reference list is presented 
in, e.g., Ref.~\cite{AGS}, and recent works on this subject are given 
in Ref.~\cite{recent}.    
By contrast, there are comparably few works concerning the {\em open\/}
string sectors of $AdS_3$ string theory, in other words,  
D-branes in the $AdS_3$
background \cite{Stan2,Stan3,BP,PR,GKSch,LOPT,Lomholt}. 

The main purpose of this paper is to propose a boundary state 
description of D-branes in (super) string theory on the $AdS_3$
background, which is known to be described by the $SL(2;\br)$ WZW model.
The method for constructing the complete basis of boundary states 
in general WZW models, called Ishibashi states, is well known
\cite{Ishibashi}.  
For this reason, it is easy to solve the general gluing conditions in any 
representation space of the current algebra. 
However, elucidating the brane interpretation of boundary states
in the $SL(2;\br)$ WZW model is still a difficult problem. 
The main difficulty originates from the fact that we have infinitely many
representations in the physical Hilbert space and thus infinitely
many Ishibashi states, which sharply contrasts with the situation in the 
$SU(2)$ WZW model. An important constraint to determine  the spectrum
of allowed boundary states is the Cardy condition \cite{Cardy},  which  
embodies the open-closed string duality. In addition to the Cardy 
condition, we stipulate the following condition: {\em Any states appearing in
both open and closed string channels of  the cylinder
amplitudes must be consistent with the requirement of the
no-ghost theorem and normalizability.}
This condition is the starting point of our investigation.
We further assume consistency with the actions of the spectral
flows, which is natural when comparing with the analysis of classical 
solutions presented in Refs.~\cite{PR,LOPT}.

This paper is organized as follows.
In section 2, we briefly review some familiar results regarding
D-branes in the $AdS_3$ background. In section 3, which is 
the main section of this paper, we study the boundary states 
constructed from on the discrete series and discuss their brane interpretation.
We further study the open string spectrum of the on-shell BPS states.  
Section 4 is devoted to discussion of several open problems.


\section{D-Branes in $AdS_3$ space}

In this section, we present a short review of the known results 
for D-branes in the $AdS_3$ space.
The $AdS_3$ space can be defined  as a hyper-surface in 4-dimensional flat
space with signature (2,2), such as
\begin{equation}
 (X_0)^2 - (X_1)^2 - (X_2)^2 + (X_3)^2 = l_{AdS}^2 ~,
\label{AdS(3)}
\end{equation} 
where $l_{AdS}$ is the radius of the $AdS_3$ space. 
We set $l_{AdS}=1$.
This space is the $SL(2;\br)$ group manifold,
and a useful parametrization is given by
\begin{equation}
 g = \left(
  \begin{array}{cc}
   X_0 + X_1 & -X_2 + X_3\\
   -X_2 - X_3 & X_0 - X_1
  \end{array}
  \right)~.
\end{equation}
Another useful parametrization is given by
\begin{equation}
g=e^{i\sigma^2\frac{t+\theta}{2}}e^{\sigma^3\rho}
e^{i\sigma^2\frac{t-\theta}{2}}~,
\end{equation}
or equivalently,
\begin{equation}
\begin{array}{l}
 X_0= \cos t \cosh \rho ~, ~~~ X_1 = \cos \theta \sinh \rho ~, \\
 X_2 = \sin \theta \sinh \rho~, ~~~ X_3= \sin t \cosh \rho~,
\end{array}
\label{global}
\end{equation}
which are called the ``global coordinates''. The $AdS_3$ metric is
written with this parametrization as
\begin{equation}
ds^2 = d\rho^2 - \cosh^2 \rho dt^2 + \sinh^2 \rho d\theta^2 ~.
\label{metric}
\end{equation}

String theory in the group manifold can be described by the WZW model.
We now consider the $SL(2;\br)_{k+2}$ WZW model (we consider the level
$k+2$ rather than the level $k$  only for convenience), 
whose action is given by 
\begin{equation}
 S = \frac{i(k+2)}{4 \pi} \int d^2 z 
\tr (g^{-1} \partial g \cdot g^{-1} \deebar g)
  + \frac{i(k+2)}{12 \pi} \int _{\cB} \tr (g^{-1} d g)^{\wedge 3}~,
\label{WZWaction}
\end{equation}
where $\cB$ is the manifold whose boundary is the world sheet.
Because we will later consider the world sheet with Lorentzian signature,
we use the light-cone coordinates  $z = e^{i(\tau + \sigma)} $ and  
$\tilde{z} = e^{i(\tau - \sigma)}$. 
This theory has the left-moving  and right-moving  conserved currents
\begin{equation}
 j(z) = - \frac{k+2}{2} \partial g \cdot g^{-1} 
= j^a T^b \eta_{ab}~,~~~
 \tilde{j} (\tilde{z}) = \frac{k+2}{2} g^{-1} \deebar g 
= \tilde{j}^a T^b \eta_{ab}~,
\label{currents 1}
\end{equation}
where we use the basis\footnote
      {With the present convention, the space-time energy operator associated 
        with the global coordinate $t$ is $j^3_0-\tilde{j}^3_0$.}
\begin{equation}
T^3 = \frac{1}{2}\sigma^2,~~~
T^{\pm} = \frac{i}{2}(\sigma^3 \pm i \sigma^1)~, 
\end{equation}
which satisfies the commutation relations
\begin{equation}
\lb T^3,~ T^{\pm}\rb = \pm T^{\pm}~, ~~~ \lb T^+,~T^-\rb =-2T^3 ~.
\end{equation}
The metric on this basis is defined by 
$\eta^{ab}= -2\tr(T^aT^b)$
(and hence, the non-zero components of $\eta^{ab}$ are $\eta^{33}=-1$, 
$\eta^{+-}=\eta^{-+}=2$), 
and we set $\eta_{ab} = (\eta^{ab})^{-1}$.

D-branes in the WZW model are described by a world sheet with a boundary, 
as established in many works \cite{dwzw}.  
In the open string description, the gluing condition of the left and 
right moving currents is generally given by
\begin{equation}
j=\om(\tilde{j})|_{z=\tilde{z}}~,
\label{glue open}
\end{equation}
where $\om$ is an automorphism of the $SL(2;\br)$ Lie algebra.
Alternatively, by exchanging the roles of the world sheet coordinates 
$\tau$ and $\sigma$, we can rewrite this condition 
in the closed string picture as 
\begin{equation}
 (j^a _n + \omega (\tilde{j})^a _{-n})\ket{B} = 0 ~,
\label{glue}
\end{equation}
where $\ket{B}$ is a boundary state that corresponds to a D-brane. 
The world-volume of such a boundary state  
can be identified with the (twined) 
conjugacy class \cite{dwzw}
\begin{equation}
 \cC ^{\omega} (h) = \{ h g \omega (h)^{-1} , ~^{\forall} h \in SL(2;\br)\}~.
\end{equation} 
If $\omega$ is an inner automorphism, we can set $\omega = {\bf 1}$ with
rotations of the currents, and the gluing condition \eqn{glue}
is thereby reduced to the simplest form,
\begin{equation}
(j^a_n+\tilde{j}^a_{-n})\ket{B}=0~.
\label{inner}
\end{equation}
We can parametrize the corresponding conjugacy classes as 
\begin{equation}
 \tr g = 2 X_0 = 2 \tilde{C}~,
\end{equation}
where $\tilde{C}$ is a constant. In this parametrization, the hyper-surface
(\ref{AdS(3)}) becomes 
\begin{equation}
 (X_3)^2 - (X_1)^2 - (X_2)^2 = 1 - \tilde{C}^2,
\end{equation}
and this equation implies  that the geometry of D-branes in the $SL(2;\br)$
group manifold is the $dS_2$ space $(\tilde{C}^2 > 1 )$ or the
hyperbolic plane ($H_2$) $(\tilde{C}^2 < 1 )$ \cite{Stan2,BP}.  
There are also  cases of  ``degenerated D-branes'',  
whose shapes are the light-cone $(\tilde{C}^2 = 1 )$ 
or the  point $(g=1)$. 
It was argued in Ref.~\cite{BP} 
using the classical analysis of the DBI action
that these D-branes are unphysical branes with 
supercritical electric fields.

If $\omega$ is an outer automorphism, we obtain a different geometry
of the brane. Let us choose $\omega$ as 
\begin{equation}
 \omega(h) = \omega h \omega^{-1} ~,~~ 
 \omega = \left(
  \begin{array}{cc}
    0 & 1\\
    1 & 0
  \end{array}
\right)~, 
\end{equation}
which defines an outer automorphism,
since $\omega$ does not belong to $SL(2;\br)$.
Then, the gluing condition becomes
\begin{equation}
\left\{
\begin{array}{rcl}
 (j^3_n - \tilde{j}^3_{-n}) \ket{B}&=& 0 \\
 (j^{\pm}_n  - \tilde{j}^{\mp} _{-n})\ket{B}&=& 0 ~, 
\end{array} \right. 
\label{outer}
\end{equation}
and the ``twined conjugacy classes'' can be expressed as
\begin{eqnarray}
 \cC ^{\omega} (h) &=& \{ h g (\omega h \omega ^{-1})^{-1} , 
   ~^{\forall} h\in SL(2;\br)\} \nn 
  &=& \{ h g \omega h^{-1} \cdot \omega^{-1}, ~^{\forall} h\in SL(2;\br)\} ~.
\end{eqnarray}
We can thus  characterize these classes by a constant $C$ as
\begin{equation}
 \tr (g \omega)  = - 2 X_2 = 2 C~.
\end{equation}
The geometry of such D-branes is given by the
hyper-surface (\ref{AdS(3)})
\begin{equation}
 (X_0)^2 + (X_3)^2 - (X_1)^2 = 1 + C^2 ~,
 \end{equation}
which is an $AdS_2$ space. 
As shown in  Ref.~\cite{BP},  such $AdS_2$-branes 
are physical D-branes with subcritical electric fields.
We concentrate on this case in this paper.

It is known that the $SL(2;\br)$ WZW model possesses symmetry
called spectral flow $U_w\otimes \tU_{\tw}$ (see, e.g.,
Ref.~\cite{HHRS}), defined by the relations
\begin{eqnarray}
&&\left\{
\begin{array}{rcl}
 U_w\,j^3 (z)\,U_w^{-1} &=& \dsp  j^3 (z) + \frac{k+2}{2}\frac{w}{z} ~,\\
 U_w\,j^{\pm} (z) U_w^{-1}&=& z^{\mp w} j^{\pm} (z) ~,
\end{array}
\right. \nonumber \\
&&
\left\{
\begin{array}{rcl} 
 \tU_{\tw}\,\tilde{j}^3 (\tilde{z})\,\tU_{\tw}^{-1} &=& 
 \dsp  \tilde{j}^3 (\tilde{z}) 
 + \frac{k+2}{2}\frac{\tw}{\tilde{z}} ~,\\
 \tU_{\tw}\,\tilde{j}^{\pm} (\tilde{z})\,\tU_{\tw}^{-1} &=& 
  \tilde{z}^{\mp \tw} 
\tilde{j}^{\pm} (\tilde{z})~,
\end{array}
\right. 
\label{spectral}
\end{eqnarray}  
which are parametrized by integers $w$ and $\tw$. 
In Ref.~\cite{MO} it is claimed that the closed string Hilbert space in the
case of the universal cover of $SL(2;\br)$
should be extended by the spectral flow with $w=-\tw$ 
(the ``winding number'').\footnote{
The time direction is uncompactified in the universal cover of
$SL(2;\br)$, and $w$ and $\tw$ must therefore be related \cite{MO}.
}
A similar claim is made in Ref.~\cite{LOPT} for the case of the open
string Hilbert space.

In our convention for currents,
the spectral flow $U_w\otimes \tU_{\tw}$  is represented by the transformation
\begin{eqnarray}
g(z,\tz) &\longmapsto &
z^{wT^3}\,g(z,\tz)\,\tz^{-\tw T^3} \nonumber\\ 
&\equiv&
e^{\frac{\tau+\sigma}{2}iw\sigma^2}\,g(z,\tz)\,
e^{-\frac{\tau-\sigma}{2}i\tw\sigma^2}~.
\label{spectral 2}
\end{eqnarray}
In this paper, we employ the single cover of $SL(2;\br)$,
which corresponds to the Wick rotation of thermal 
$AdS_3$ space. This is because we are interested in 
the open-closed string duality, 
in which the roles of the world sheet coordinates $\tau$ and 
$\sigma$ are exchanged, and hence we need to consider 
the winding sectors along not only the space-like circle but also 
the time-like circle in some situations.
In this case we can a priori choose the left-moving 
and right-moving windings $w$ and $\tw$ independently.
However, the requirement of consistency with the gluing condition yields 
some constraints on the windings $w$ and $\tw$. 
For the gluing condition \eqn{inner}, the allowed spectral flows 
should have the forms $U_w\otimes \tU_{-w}$, in other words, 
\begin{equation}
\left\{
\begin{array}{lll}
 t &\mapsto &t+w\tau \\
 \theta &\mapsto&\theta + w\sigma \\
 \rho &\mapsto&\rho ~,
\end{array}
\right.
\label{spectral inner}
\end{equation}
in terms of the global coordinates \eqn{global}
in the closed string channel.    
The spectral flows allowed  for \eqn{outer}
are $U_w\otimes\tU_w$, in other words, 
\begin{equation}
\left\{
\begin{array}{lll}
 t &\mapsto &t+w\sigma \\
 \theta &\mapsto&\theta + w\tau \\
 \rho &\mapsto&\rho ~,
\end{array}
\right. 
\label{spectral outer}
\end{equation}
which generates non-trivial winding sectors along the time-like circle.


\section{Boundary states in string theory on $AdS_3$}

In this section we attempt to construct consistent 
boundary states describing the D-branes in the $AdS_3$ space. 
As we stated in the Introduction, our  main criterion is 
the open-closed  string duality with the requirement that the spectra in 
both open and closed string channels be compatible 
with unitarity and normalizability and moreover the spectral flow
symmetry. Here, we treat only the states belonging 
to the principal discrete series (short string sector\footnote
    {In this paper we use the terminology  ``short string'' and 
     ``long string'',  following  Ref.~\cite{MO}. Here, a short 
     string is an excitation corresponding to the discrete series
     (with arbitrary windings $w$), and a long string corresponds 
     to the continuous series (with non-zero windings). 
      We call the sectors 
      with non-zero $w$ the ``winding strings'' 
     or ``circular strings''. }) 
and leave the case of the principal continuous series to future works.

To fix a specific background, let us consider the  superstring 
vacua $AdS_3 \times S^3 \times T^4$, 
which is the most familiar example (see, e.g. Ref.~\cite{GKS}), although 
we shall mainly focus on the bosonic sector. 
The $AdS_3$ sector is described by the $SL(2;\br)_{k+2}$ super WZW model 
 (where $k+2$ is the level of bosonic current) and 
the $S^3$ sector is described by the $SU(2)_{k-2}$ super WZW model.
We here assume $k\in \bz$ and $k>2$.

In string theory on Lorentzian $AdS_3$  
it is known \cite{MO} that the physical Hilbert space should
be constructed using the representation spaces of discrete series
with the constraints of the no-ghost theorem \cite{sl2,EGP} and
normalizability, and also using the continuous series. 
We should incorporate the degrees of freedom of spectral flow for
both of these representations, as discussed in Ref.~\cite{MO}.

\subsection{Open-closed duality on the Lorentzian world sheet}

Before presenting our main analysis, we would like to 
make a few comments about the open-closed duality 
on the world sheet with the {\em Lorentzian signature},
which we use in subsequent analysis.

In the case of a Euclidean world sheet, in the standard argument,
the open-closed duality for the cylinder amplitude is expressed 
as the relation\footnote
   {Throughout this paper we denote the moduli for the open string channel
    by $\tau$ and $z$ and those for the closed string channel by 
    $\displaystyle \tilde{\tau}\left(\equiv -\frac{1}{\tau}\right)$ and 
    $\displaystyle \tilde{z}\left(\equiv \frac{z}{\tau}\right)$. We also write 
    $\tilde{q}\equiv e^{2\pi i \tilde{\tau}}$, $\tilde{y}\equiv 
    e^{2\pi i \tilde{z}}$, and so on.}
\begin{equation}
\int_0^{\infty} dT^{(c)} \, \eta(\ttau)^2\, Z_{\msc{closed}}(\ttau)
=\int_0^{\infty} \frac{dT^{(o)}}{T^{(o)}}\, \eta(\tau)^2
\, Z_{\msc{open}}(\tau) ~,
\label{open-closed}
\end{equation}
where $\tau = iT^{(o)}$ denotes the open string modulus and 
 $\dsp  \ttau \equiv -\frac{1}{\tau} = iT^{(c)} $ denotes the closed 
string modulus. The factors of $\eta$-functions are the contributions 
from $bc$-ghosts.

From the simple identities
\begin{equation} 
 \int_0^{\infty} dT^{(c)}=
 \int_0^{\infty} \frac{dT^{(o)}}{T^{(o)\, 2}}~,~ 
\eta(\ttau)^2= -i\tau\eta(\tau)^2= T^{(o)}\eta(\tau)^2 ~,
\end{equation}
we obtain 
\begin{equation}
 Z_{\msc{closed}}(\ttau) =  Z_{\msc{open}}(\tau) ~,
\label{open-closed E}
\end{equation} 
which is the standard  statement of the open-closed duality.

In the case of the Lorentzian world sheet,
one must regard the moduli $\tau$ and
$ \ttau \left(\equiv -\frac{1}{\tau}\right)$ 
as real numbers. In  this paper we adopt the convention
$\tau = -T^{(o)} (<0)$ and
$\dsp \ttau 
= T^{(c)}(>0)$
(such that $ T^{(o)}= \frac{1}{T^{(c)}}$ holds as before).
Then, \eqn{open-closed} reduces to the relation
\begin{equation}
 Z_{\msc{closed}}(\ttau) = -i  Z_{\msc{open}}(\tau) ~,
\label{open-closed L}
\end{equation}
instead of \eqn{open-closed E}. 
Clearly, the same formula holds also for the superstring case.


\subsection{Boundary states based on discrete series}

We employ the  (bosonic) $SL(2;\br)_{k+2}$ WZW model.    
Let $\hat{\cD}^{\pm}_l$ be the representation space of  
discrete series
(where $+$ corresponds to the  lowest weight representation, and $-$
corresponds to the  highest weight 
representation).\footnote{
   With our convention, the conformal weight of zero-mode states
    is given by $\dsp h= -\frac{l(l+2)}{4k}$, and the $j^3_0$-spectrum of
    the zero-mode states is $\dsp j^3_0 =
    \pm\left(\frac{l}{2}+n+1\right)$, $(n\in \bz_{\geq 0})$, 
    for $\hat{\cD}^{\pm}_l$ ($l>-2$). 
    The double-sided  representations (``degenerate 
     representations'') have the zero-mode spectra $\dsp j^3_0= \frac{l}{2},
    \frac{l-2}{2}, \ldots , -\frac{l}{2}$ ($l\in \bz_{\geq 0}$), 
    which are   natural analogs of the unitary representations 
    of $SU(2)$.  
    On the other hand,  the principal continuous series  
    $\hat{\cC}_{\la,\al}$ ($\la \in \br$, $0\leq \al <1$), which
    corresponds to the branch $l=-1+ 2i \la$,  i.e., 
    $\dsp h= \frac{1}{k}\left(\la^2+\frac{1}{4}\right)$ and has
     the zero-mode spectrum $j^3_0 = \al +n$, ($n\in \bz$).}
We also express the representation  transformed by the spectral flow as
$\hat{\cD}^{\pm\,(w)}_l$,  as in Ref.~\cite{MO}. 
The spectral flow parameter $w$ ($=\tw$ for the $AdS_2$-brane) is
introduced as \eqn{spectral} in our convention. 
The unitarity and normalizability lead to the constraints 
$-1 < l < k-1$  for the $l$-quantum number \cite{MO,EGP,GK}.\footnote{
The requirement of the no-ghost theorem is $-2 < l < k$, and the
requirement of normalizability is more severe. 
The wave function of discrete series behaves as $\sim e^{-(l+1)\rho}$, as we
demonstrate below, and hence $L^2$ normalizability requires $l >-1$.  
Moreover, we assume that the spectral flow is the symmetry of the model,
and thus the states characterized by $l$ and $k -l - 2$ are related by
this symmetry, and the requirement becomes $-1 < l < k-1$.
}
Since we are here considering a single cover of the $AdS_3$ space,
the allowed values of $l$ are $l=0,1, \ldots, k-2$. 
Quite interestingly, this range is the same as 
that  for the integral representations of $SU(2)_{k-2}$.

The character with respect to the unflowed representation
$\hat{\cD}^{\pm}_l$ is given by
\begin{equation}
\chi^{\pm}_l(\tau,z) 
\equiv \tr_{\hat{\cD}^{\pm}_l} q^{L_0-\frac{c}{24}}y^{j^3_0} = 
\frac{q^{- \frac{(l+1)^2}{4k}} 
y^{ \pm \frac{l+1}{2}}}
{- i\th_1(\tau,\pm z)} ~,
\label{character -1}
\end{equation}
where $q\equiv e^{2\pi i \tau}$, $y\equiv e^{2\pi i z}$.
The character with respect to the flowed representation
$\hat{\cD}^{\pm\,(w)}_l$ is given in, for example, Refs.~\cite{HHRS,MO}.
By using the relation between the unflowed and flowed currents
(\ref{spectral}), the character with respect to $\hat{\cD}^{\pm\,(w)}_l$
can be rewritten in the terms of the character 
(\ref{character -1}), and we obtain
\begin{equation}
\chi^{\pm\, (w)}_l(\tau,z) 
\equiv \tr_{\hat{\cD}^{\pm\,(w)}_l} q^{L_0-\frac{c}{24}}y^{j^3_0} = 
(-1)^w \frac{q^{- \frac{(l+1\mp kw)^2}{4k}} 
y^{ \pm \frac{l+1 \mp kw}{2}}}
{- i\th_1(\tau,\pm z)} ~.
\label{character 0}
\end{equation}

Let us start with the Ishibashi state $\ket{l, \pm}_I$ 
obtained with the representation $\hat{\cD}^{\pm}_l$ defined 
by the gluing condition \eqn{outer} (with $a=0$)
describing the $AdS_2$-branes. 
Let $\cB (\hat{\cD}^{\pm}_l)$ be an  orthonormal basis of 
$\hat{\cD}^{\pm}_l$ composed of the eigenstates of $j^3_0$.
We choose the phases of each bases $\bv \in \cB
(\hat{\cD}^{\pm}_l) $ so that the matrix elements of all the currents
$j^3_n$, $j^{\pm}_n$ are real numbers.  
More precisely, we assume that 
$\bra{\bv_1} j^3_n \ket{\bv_2}= \bra{\bv_2} j^3_{-n} \ket{\bv_1}$,
$\bra{\bv_1} j^{\pm}_n \ket{\bv_2}= \bra{\bv_2} j^{\mp}_{-n}
\ket{\bv_1}$ for arbitrary $\bv_1, \, \bv_2 \in \cB (\hat{\cD}^{\pm}_l)$. 
$\cB (\hat{\cD}^{\pm}_l)$ is not uniquely defined 
even under this requirement. However, this ambiguity clearly causes no
problem our analysis.
Further, we denote  the signature of norm $\langle \bv | \bv
\rangle$ as $\ep_{\bsv}$ for each $\bv \in \cB (\hat{\cD}^{\pm}_l)$.  
(Recall that $\hat{\cD}^{\pm}_l$ is not a unitary representation 
of the affine algebra $\widehat{SL}(2;\br)_{k+2}$.)

With these preliminaries, we can explicitly write down 
the Ishibashi state that satisfies the gluing condition (\ref{outer}),
\begin{equation}
\ket{l, \pm}_I = \sum_{\bsv \in\cB(\hat{\cD}^{\pm}_l)}\, \ep_{\bsv} 
\ket{\bv}\otimes \widetilde{\ket{\bv}}~.
\label{Ishibashi discrete}
\end{equation}
The gluing condition \eqn{inner} (for the conjugacy class with no twist) 
is also easily solved, and we find
\begin{equation}
\ket{l, \pm}'_I = \sum_{\bsv \in \cB(\hat{\cD}^{\pm}_l)}\, 
\ep_{\bsv}\ket{\bv}\otimes T\widetilde{\ket{\bv}}~,
\label{Ishibashi discrete 2}
\end{equation}
where $T$ denotes the isomorphism 
$T~:~\hat{\cD}^{\pm}_l~\stackrel{\cong}{\rightarrow}~\hat{\cD}^{\mp}_l$,
such that $T$ maps the lowest (highest) weight vector 
in $\hat{\cD}^+_l$ ($\hat{\cD}^-_l$) to the highest (lowest)
weight vector in $\hat{\cD}^-_l$ ($\hat{\cD}^+_l$), and satisfies
\begin{eqnarray}
T\, \tilde{j}^3_n\,T &=& - \tilde{j}^3_n ~, \nonumber \\
T\, \tilde{j}^{\pm}_n\,T &=& - \tilde{j}^{\mp}_n ~.
\end{eqnarray}
However, since the brane configuration described by \eqn{inner}
is known to be unphysical \cite{BP},
we concentrate on the $AdS_2$-brane cases \eqn{outer}.

As pointed out above, the spectral flow compatible with 
the gluing condition \eqn{outer} is of the type \eqn{spectral outer}, 
and we can similarly obtain the Ishibashi states for 
the flowed representations:
\begin{equation}
\ket{l, w, \pm}_I = \sum_{\bsv \in \cB(\hat{\cD}^{\pm\,(w)}_l)}\, 
\ep_{\bsv} \ket{\bv}\otimes
\widetilde{\ket{\bv}}~.
\label{Ishibashi discrete flow}
\end{equation}
These states are characterized by the cylinder amplitudes
\begin{equation}
{}_I\bra{l,\pm, w}\tilde{q}^{H^{(c)}}\tilde{y}^{j^3_0}
\ket{l',\pm, w'}_I = \delta_{ll'}  \delta_{ww'}\,
\chi^{\pm\, (w)}_l(\tilde{\tau},\tilde{z})~, 
\end{equation}
where $H^{(c)} = \frac{1}{2} (L_0 + \tilde{L}_0 - \frac{c}{12})$ 
and $\chi^{\pm\, (w)}_l(\tilde{\tau},\tilde{z})$ is defined 
in \eqn{character 0}.

Now, we consider the following boundary states, which will be
used as the building blocks of our Cardy states:
\begin{eqnarray}
\ket{l,w}_I  &=& \ket{l,w,+}_I + \ket{l,-w,-}_I \nn
&\equiv& \ket{l,w,+}_I + \ket{k-2-l,-w+1,+}_I ~.
\label{bb}
\end{eqnarray}
(These states are also used in Ref.~\cite{Yamaguchi}.) 
For these states, we have 
\begin{eqnarray}
{}_I\bra{l,w} \tilde{q}^{H^{(c)}}\tilde{y}^{j^3_0} \ket{l',w'}_I
&=& \delta_{ll'}\delta_{ww'} (-1)^w \left\lb
\chi^{+ \, (w)}_l(\tilde{\tau},\tilde{z}) 
+ \chi^{- \, (-w)}_l(\tilde{\tau},\tilde{z}) \right\rb  \nn
&=& \delta_{ll'}\delta_{ww'} (-1)^w
\left\lb \frac{q^{-\frac{(l+1-kw)^2}{4k}} 
y^{ \frac{l+1-kw}{2}}}
{- i\th_1(\tilde{\tau}, \tilde{z})} 
+ \frac{q^{-\frac{(l+1-kw)^2}{4k}} 
y^{-\frac{l+1-kw}{2} }}
{i\th_1(\tilde{\tau}, \tilde{z})} \right\rb \nn
&\equiv&  \delta_{ll'}\delta_{ww'} \,  
\chi^{(w)}_l (\tilde{\tau},\tilde{z})~. 
\label{bb amplitude}
\end{eqnarray}
The right-hand side of the above identity \eqn{bb amplitude}
is well-defined in the limit $\tilde{z}\,\rightarrow\, 0$,
where it becomes\footnote{
It is interesting that the character $\chi_l^{(w=0)}(\tau, z) $
is formally equal to (the opposite sign of) 
the character of the degenerate representation treated in Ref.~\cite{GKSch}. }
\begin{eqnarray}
{}_I\bra{l,w} \tilde{q}^{H^{(c)}} \ket{l',w'}_I
&=& \delta_{ll'}\delta_{ww'} (-1)^w \left\lb 
-(l+1-kw) \frac{\tilde{q}^{-\frac{(l+1-kw)^2}{4k}}}{\eta(\tilde{\tau})^3}
\right\rb  \nn
&\equiv & \delta_{ll'}\delta_{ww'} \,  \chi^{(w)}_l (\tilde{\tau})~.
\label{bb amplitude 2}
\end{eqnarray}
In fact, the existence of $z$ makes the character (\ref{character 0})
convergent, 
and if we take the naive limit $z \to 0$, this character would be
divergent. However by using the combination (\ref{bb}), we can remove this
divergence. This divergence is due to the infinite
dimensional representation of $SL(2;\br)$ and this type of divergence can be
removed using the usual regularization, like that of the $\zeta$ function.
The combination (\ref{bb}) cancels the divergence and also removes
the regulator. (The detailed discussion about the subtlety of this
character is given in Ref.~\cite{HR}.)

It is worthwhile to note that the summation 
of the characters
$ \sum_{w\in 2\bsz}\chi^{(w)}_l (\tau, z)$
is formally quite similar to the well-known character formula 
of $\widehat{SU}(2)_{k-2}$,  
as was pointed out in Ref.~\cite{HHRS}. 
In fact, we have\footnote
 {The range of summation $w \in 2 \bz$ in $\chi^{SL(2)}_l(\tau,z)$
indicates that we sum all the windings $w$,
since we use the combination (\ref{bb}) as the building blocks.}
\begin{equation}
\chi_l^{SL(2)} (\tau,z) \equiv  \sum_{w\in 2\bsz} \chi^{(w)}_l(\tau,z)
=  \frac{\Th{-(l+1)}{-k} (\tau ,z)-
 \Th{l+1}{-k} (\tau,z)}
 {i\th_1(\tau,z)} ~,
\label{character SL2}
\end{equation}
and for $\widehat{SU}(2)_{k-2}$, the character is written 
\begin{equation}
\chi^{SU(2)}_l(\tau,z) = \frac{\Th{l+1}{k} (\tau,z)-
 \Th{-(l+1)}{k} (\tau,z)}
 {i\th_1(\tau,z)}~.
\label{character SU2}
\end{equation}
This fact leads to the good modular property of the character 
$\chi^{SL(2)}_l$,
\begin{equation}
\chi^{SL(2)}_L(-\frac{1}{\tau},\frac{z}{\tau}) = i  \sum_{l=0}^{k-2}
   S^{(k-2)}_{Ll} \chi^{SL(2)}_l(\tau,z)~,
\label{modular SL2}
\end{equation}
where 
\begin{equation}
 S^{(k-2)}_{Ll}\equiv\sqrt{\frac{2}{k}}
\sin \left(\pi \frac{(L+1)(l+1)}{k}\right) 
\end{equation}
is the well-known matrix of the modular transformation 
for $\widehat{SU}(2)_{k-2}$.

Of course, the power series of \eqn{character SL2}
is divergent for the usual range of the modulus $\tau$ 
(i.e.  $\mbox{Im}\, \tau >0$), 
since they include the negative level theta functions. 
Therefore, we must here adopt the Lorentzian signature
on the world sheet; that is,
we must regard $\tau$  as a real number, as in the calculation 
of partition function in the appendix of Ref.~\cite{MO}.
The extra factor $i$ on the RHS of \eqn{modular SL2} reflects this fact,
and its existence matches the formula of the open-closed duality 
in the Lorentzian world sheet \eqn{open-closed L}.
One may suppose that we still have a subtlety in  \eqn{modular SL2} 
(and also in the similar formulas given below),  
due to the fact that the power series does not absolutely converge. 
However, we can give this a strict meaning with the help of 
generalized functions. We here simply present such formal identities  
to avoid notational complexities. Their mathematically rigorous  
treatment is given in Appendix B.


Now, we would like to construct the suitable Cardy states from 
the Ishibashi states \eqn{bb}.
We start with the ansatz
\begin{equation}
\ket{a}_C = \sum_{l=0}^{k-2}\, 
\sum_{w\in 2\bsz} \Psi_a(l,w)\ket{l,w}_I~,
\label{ansatz}
\end{equation}
where the index $a$ runs over the set of allowed Cardy states, which
should be fixed later.
It should be emphasized that the summation over the winding $w$
is needed for the good modular property.
From the unitarity bound for the closed string spectrum, 
we must take the range of the $l$-summation as $0\leq l \leq k-2$.

We first note the modular property of the character 
$\chi_l^{(w)}(\tau)$,
\begin{equation}
\sum_{w\in 2\bsz} \sum_{l=0}^{k-2}\, \sqrt{\frac{2}{k}}
\sin\left(\pi \frac{(L+1)(l+1-kw)}{k}\right)\, \chi_l^{(w)}(\ttau)
= -i \sum_{W\in 2\bsz} \chi_L^{(W)}(\tau) ~,
\label{modular disc}
\end{equation}
which can be derived by making use of the Poisson resummation formula.
We can here assume $-1 < L < k-1$ without loss of generality,
because $L$ appears only in the combination $L'=L+1- kW$
in the character $\chi_L^{(W)}(\tau) $.
Moreover, since the time direction is compactified,
the character in the open string channel $\chi_L^{(W)}(\tau)$
must again be characterized by the discrete quantum number $L$. 
Therefore, we can assume $L=0,1,\ldots, k-2$, and 
the Cardy condition for the ansatz \eqn{ansatz}
can  be written 
\begin{equation}
\Psi_a^*(l,w)\Psi_b(l,w) = \sum_{L=0}^{k-2}\, N^{(o)}_{ab}(L)
\sqrt{\frac{2}{k}}\sin\left(\frac{\pi (L+1)(l+1)}{k}\right) ~,
\label{Cardy cond}
\end{equation}
where $N^{(o)}_{ab}(L) \in \bz_{\geq 0}$.
This condition is formally the same as that of
$\widehat{SU}(2)_{k-2}$, and the solution is well known \cite{Cardy}.
Assuming the diagonal modular invariant, the solution is given by
\begin{eqnarray}
\ket{L}_C &=& \sum_{l=0}^{k-2} \sum_{w\in 2 \bsz}\Psi_{L}(l,w)
\ket{l,w}_I ,\nn
 \Psi_{L}(l,w) &=& 
 \frac{S^{(k-2)}_{Ll}}{\sqrt{S^{(k-2)}_{0l}}} \nn
&=& 
\left(\frac{2}{k}\right)^{1/4} \frac{\sin\left(\frac{\pi (L+1)(l+1)}{k}\right)}
{\sqrt{\sin\left(\frac{\pi (l+1)}{k}\right)}}~.
\label{Cardy solution}
\end{eqnarray}

Using the identity
\begin{equation}
\frac{S^{(k-2)}_{L_1l}S^{(k-2)}_{L_2l}}{S^{(k-2)}_{0l}}
= \sum_{L=0}^{k-2}N_{L_1,L_2}^L    S^{(k-2)}_{Ll}~,
\label{Verlinde}
\end{equation}
where $N_{L_1L_2}^L$ denotes the fusion matrix of $\widehat{SU}(2)_{k-2}$
\begin{equation}
N_{L_1L_2}^L =\left\{
\begin{array}{ll}
 1 & ~~ |L_1-L_2| \leq L \leq \mbox{min}(L_1+L_2,\, 2(k-2)-L_1-L_2) \\
   &  \hspace{2cm}    \mbox{and} ~ L\equiv |L_1-L_2| ~\mod ~2 \\
  0 & ~~  \mbox{otherwise} ~,
\end{array}
\right.
\end{equation}
we obtain, as in the $SU(2)$ case, 
\begin{equation}
{}_C\bra{L_1} \tilde{q}^{H^{(c)}} \ket{L_2}_C = - i 
\sum_{L=0}^{k-2} N_{L_1L_2}^L \chi^{SL(2)}_L(\tau)~,
\label{Cardy relation}
\end{equation}
where 
\begin{equation}
\chi^{SL(2)}_L(\tau) \equiv \sum_{W\in 2\bsz} \chi^{(W)}_L(\tau)
\equiv  \sum_{W\in 2\bsz}\left\lb 
-(L+1-kW)\frac{q^{-\frac{(L+1-kW)^2}{4k}}}{\eta(\tau)^3}
\right\rb ~,
\label{character SL2 2}
\end{equation}
as defined above.
The identity \eqn{Cardy relation} corresponds to the relation of  
the open-closed duality \eqn{open-closed L} in which we are interested.

As we stated previously, the important criterion 
for the Cardy states is the requirement that 
the spectrum in both open and closed string channels 
be consistent with the unitarity bound. Moreover, it is quite natural 
to require that the density of states in the open string channel 
be invariant under the spectral flow, as claimed in Ref.~\cite{LOPT}
based on the analysis of the classical open string solutions.
Equation \eqn{Cardy relation} actually possesses these properties, and
hence \eqn{Cardy solution} is regarded as the desired solution 
of the Cardy condition.


\subsection{Identification of $AdS_2$-branes with weak electric
fields}

It is discussed  in Ref.~\cite{BP} that the physical D-brane (D-string)
in the $AdS_3$ background should be wrapped on the twined conjugacy class,
which has the structure of the $AdS_2$ space.
Such an $AdS_2$-brane can be expressed as the following simple equation
with respect to the global coordinates of $AdS_3$ $(t,\theta, \rho)$
\eqn{global}:
\begin{equation}
\sinh \rho \sin \theta = \mbox{const.} \equiv \sinh \psi_0~.
\label{AdS2 brane}
\end{equation}
Here $\psi_0$ parameterizes the location of the $AdS_2$-brane
and corresponds physically to the strength of the electric field 
on that brane. In fact, we can readily find from \eqn{AdS2 brane}
that $\rho \geq \rho_{\msc{min}}\equiv \psi_0 $,
and thus $\psi_0$ parametrizes the point nearest to the center of
the $AdS_3$ space,  on  the $AdS_2$-brane ``bent'' by the electric field.

We  now  attempt to identify the Cardy states \eqn{Cardy solution}
with the $AdS_2$-branes. Our arguments are summarized as follows:
\begin{enumerate}
 \item From the construction and evaluation of the  cylinder
amplitudes \eqn{Cardy relation},
it is clear that our Cardy states \eqn{Cardy solution}
can interact only with the short string sectors in  
both open and closed string channels.  
Moreover, the fact that we are now taking   
the discrete values of $l$ implies that only the string modes 
propagating within the range $\rho \lsim 1$ 
(in the unit of the $AdS_3$ scale $l_{AdS}$)
can interact with the Cardy states \eqn{Cardy solution}.
In fact, the wave function of the string states corresponding to
$l$ is known to behave\footnote{
The wave function of the discrete series behaves as 
$\sim e^{- (l+2) \rho} / g_s$, where $g_s$ is the string coupling.
Now it is given by $g_s = e^{- \rho}$. (See, for example, Ref.~\cite{SW}.)
}  as $\sim e^{-(l+1)\rho}$ with respect to the 
``radial coordinate'' $\rho$ (where $l=-1$ corresponds to the 
Breitenlohner-Freedman bound \cite{BrF}).  
Therefore, since we are now working with
$l=0, 1, \ldots, k-2$, the corresponding wave functions  
exponentially damp at the length scale $\rho \sim 1$. 
\item 
At least in the large $k$ limit, our cylinder amplitudes  
must be interpreted as the summation of classical open string 
solutions. 
We have non-trivial winding sectors generated 
by spectral flows of the type $t~\rightarrow~t+w\tau$,
$\theta~\rightarrow~\theta+w\sigma$.   
Such classical solutions have been studied in several recent
works \cite{PR,LOPT}.   In particular, explicit forms of the 
classical open short string solutions connecting two $AdS_2$-branes 
labeled $\psi_1$ and $\psi_2$ are given in Ref.~\cite{LOPT} (up to the
degrees of freedom of  $SL(2;\br)$ isometry). These are 
\begin{equation}
\left\{
\begin{array}{l}
 t=(\al+w)\tau ~,\\
 \theta= (\al+w)\sigma+\theta_0 ~,\\
 \rho=\rho_0 ~,
\end{array}
\right.
\label{classical solution 1}
\end{equation}
with even winding $w\in 2\bz$, and the parameters $\al$
$(0\leq \al \leq 1)$, $\theta_0$ $(0\leq \theta_0\leq 2\pi)$, 
$\rho_0(\geq 0)$ must satisfy
\begin{equation}
\left\{
\begin{array}{l}
  \sinh \rho_0 \sin \theta_0 =\sinh \psi_1 ~,\\
  \sinh \rho_0 \sin (\al \pi+ \theta_0) =\sinh \psi_2 ~.
\end{array}
\right.
\end{equation}
For odd winding $w\in 2\bz+1$, we similarly have 
\begin{equation}
\left\{
\begin{array}{l}
 t=(1-\al+w)\tau ~,\\
 \theta= (1-\al+w)\sigma+\pi-\theta_0 ~,\\
 \rho=\rho_0 ~,
\end{array}
\right.
\label{classical solution 2}
\end{equation}
where $\al,~ \theta_0,~ \rho_0$ are the same as above.

\hspace{0.575cm}
The building blocks of the open string amplitudes
\eqn{Cardy relation} are the characters 
$\chi^{SL(2)}_L(\tau)$ \eqn{character SL2 2}, which can be rewritten as
\begin{eqnarray}
\chi^{SL(2)}_L(\tau) &=& \lim_{z\rightarrow 0}\left\lb 
-\sum_{w\in 2\bsz}\frac{q^{-\frac{(L+1-kw)^2}{4k}} 
y^{\frac{L+1}{2}-\frac{kw}{2}} }
{i\th_1(\tau,z)} \right. \nn
 && \left. \hspace{1cm}+\sum_{w\in 2\bsz+1}\frac{q^{-\frac{(k-L-1-kw)^2}{4k}} 
y^{\frac{k-L-1}{2}-\frac{kw}{2}} }
{i\th_1(\tau,z)}
\right\rb ~.
\label{character SL2 3}
\end{eqnarray}
It is quite natural to interpret the zero-mode parts of 
these amplitudes  as the summation over the classical solutions.
In fact, one can easily find that the first term 
and the second term in \eqn{character SL2 3}
correspond nicely to the classical solutions
\eqn{classical solution 1} and \eqn{classical solution 2}, respectively,
since the energy parameter $\al$ is quantized as $n/(k+2)$, 
because of the time-like compactification, and hence identified with 
$(L+1)/k$ in the large $k$ limit. 
(The classical solution
\eqn{classical solution 1} possesses the classical conformal weight 
$-\frac{k+2}{4} \al^2 \approx -\frac{k}{4} \al^2 $, 
which should correspond to the quantum value
$-\frac{(L+1)^2}{4k}+ \frac{1}{4k}$.)

\hspace{0.575cm}
We can also check the consistency of spectral flows 
in the open and closed string channels. 
The spectral flows that generate the classical solutions 
\eqn{classical solution 1} and \eqn{classical solution 2} 
are equivalent to $U_w\otimes \tU_w$ and
\eqn{spectral outer} in the closed string channel,
after exchanging the roles of $\tau$ and $\sigma$. 
These are  in fact compatible with the gluing condition \eqn{outer},
as we noted above. 

\hspace{0.575cm}
If we consider the $dS_2$-branes instead of the $AdS_2$-branes,
classical open string solutions with non-trivial winding numbers
are generated by the spectral flows: $t~\rightarrow~t+w\sigma$,
$\theta~\rightarrow~\theta+w\tau$. These are equivalent to 
\eqn{spectral inner} in the closed string channel and compatible 
with the gluing condition \eqn{inner}.
\item Recalling the previous results obtained for the
$SU(2)$ WZW model, it seems plausible 
to relate the labels of the Cardy states $L_1$ and $L_2$
with the parameters of the brane positions $\psi_1$ and $\psi_2$.
However, we would immediately face an apparent contradiction. 
In \eqn{Cardy relation},
the $L$-value appearing in the open string channel
has upper and lower bounds. However, the corresponding 
parameter $\al$ in the classical solutions
should not have such bounds depending on the brane positions 
$\psi_1$ and $\psi_2$, as discussed in Ref.~\cite{LOPT}. 
How should we resolve this apparent contradiction? 
Recall the fact that our Cardy states 
only include the excitations of short strings propagating
within  the finite domain $\rho \lsim 1$.
This implies that the classical solutions \eqn{classical solution 1} and
\eqn{classical solution 2} we should compare with the
open string spectrum 
have to satisfy the constraints $\rho_0 \lsim 1$. 
We can hence expect the upper and lower bounds for $\al$ in 
\eqn{classical solution 1} and \eqn{classical solution 2}.

\hspace{0.575cm}
Interestingly, assuming the identification 
$ \psi_i \approx \frac{\pi}{2}\left(1-\frac{2L_i}{k}\right) $
for sufficiently small $|\psi_i|$ (weak electric field) 
and requiring $\sinh \rho_0 \lsim \sinh 1 \sim 1$, 
we can show by a simple geometrical argument  that 
\begin{equation}
|L_1-L_2| \lsim k \al \lsim \min (L_1+L_2, 2k-L_1-L_2)~.
\label{bound al}
\end{equation}
This relation nicely reproduces the quantum truncation 
appearing in the cylinder amplitude \eqn{Cardy relation}.
Based on these consideration, we assert
that the quantum number $L$ that labels 
the Cardy states should correspond to the parameter $\psi_0$
parametrizing the locations of the $AdS_2$-brane. Our above observation
supports this assertion, at least in cases in which $|\psi_0| \ll 1$.

\hspace{0.575cm}
Of course, since $L$ can only take a value within the range 
$L=0, 1, \ldots, k-2$, the number of allowed branes
should be finite
in our quantum analysis. 
(In particular, 
our Cardy states cannot describe the $AdS_2$-brane with 
the strong electric field whose entire world-volume is located outside 
the $AdS$ radius.) This situation is quite similar to that
in the $SU(2)$ case,  and has its origin in the existence
of the unitarity bound. 

\hspace{0.575cm}
We also point out that the $\bz_2$ symmetry 
$\psi_i \, \rightarrow \, -\psi_i$
corresponds to $L_i \, \rightarrow \, k-2-L_i $,
and the cylinder amplitudes \eqn{Cardy relation} possess 
this symmetry, as expected. 
The existence of quantum truncation 
is again necessary for this to hold.

\end{enumerate}

\subsection{Space-time chiral primaries in the open string spectrum}

Now, let us turn our attention to the superstring on the background
$AdS_3 \times S^3 \times T^4$. Based on our analysis in the previous 
subsections  and incorporating the $SU(2)$ WZW model and free fermions, 
we can determine  the open string spectrum in this superstring theory. 
We denote the free fermions along the $SL(2;\br)$ directions as 
$\psi^3$ and $\psi^{\pm}$ and those of $SU(2)$ as $\chi^3$ and $\chi^{\pm}$.

We focus on the special class of physical states - ``space-time (anti) chiral 
primary states'', which are inevitably $1/2$ BPS states.
For the closed string sector, such BPS states 
(or the corresponding vertex operators) are 
important in the context of $AdS_3/CFT_2$ correspondence and are 
investigated in Refs.~\cite{KLL,HS,HHS,AGS}.  
We show that the BPS 
$AdS_2 \times S^2$-brane whose $SL(2;\br)$ sector 
is described by our Cardy state \eqn{Cardy solution}
contains an infinite number of such excitations compatible with 
the spectral flows.

The Cardy state of the total system should have 
the structure
$$
\ket{L,L',\ldots}_C \equiv \ket{L}^{SU(2)}_C\otimes \ket{L'}^{SL(2)}_C
\otimes \cdots ~,
$$
where $\ket{L}^{SU(2)}_C$ is the Cardy state of the $SU(2)$ sector,
$\ket{L'}^{SL(2)}_C$ denotes the Cardy state defined in \eqn{Cardy
solution}, and $\cdots$ represents the contributions from other sectors
(free fermions and the $T^4$ sector). Considering a specific cylinder
amplitude with respect to such a Cardy state, we typically obtain 
the amplitude as 
$$
\dsp \sum_{L,L'}N^L_{L_1L_2}N^{L'}_{L'_1L'_2}\chi^{SU(2)}_L(\tau)
\chi^{SL(2)}_{L'}(\tau)\cdots ~,
$$ 
where $N^{L}_{L_1L_2}$ and $N^{L'}_{L'_1L'_2}$ denote again
the fusion coefficients of $\widehat{SU}(2)_{k-2}$. 
The most important fact for our later discussion is that 
the character $\chi^{SL(2)}_L(\tau)$ \eqn{character SL2 2}
contains the contributions from the infinitely many sectors with
non-trivial winding numbers.
This leads us  to a rich structure of the spectrum of BPS states.

Now, let us begin the analysis of on-shell BPS states.
We only consider the NS sector here and work with 
the $(-1)$-picture. We also ignore the excitations along 
the $T^4$ direction for simplicity. 
First, we focus on the sector of primary states 
(in the usual sense of the world sheet) with $w=0$, 
and consider the physical states including only one oscillator 
of free fermions (``level one states'').
The on-shell condition stipulates that the spin of the $SL(2;\br)$ sector 
must be equal to that of $SU(2)$. This means that we must look for the BRST
invariant states within the Hilbert space,
$\hat{\cL}_L \otimes \hat{\cD}^+_L (\otimes ~ \mbox{Hilbert space of 
the free fermions})$ or $\hat{\cL}_L \otimes \hat{\cD}^-_L$, 
where $\hat{\cL}_L$ represents the integrable representation with the spin
$L/2$ of $\widehat{SU}(2)_{k-2}$.  
For example, in the Hilbert space $\hat{\cL}_L\otimes \hat{\cD}^+_L$,
generic physical states (of level one) should have the form
\begin{equation}
\begin{array}{l}
\dsp \sum_{M,M',A}
  \al_{M,M',A}\ket{L,M}^{SU(2)} \otimes \ket{L,M'}^{SL(2)} \otimes
  \psi^A_{-1/2} \ket{0}_f \otimes ce^{-\phi}\ket{0}_{\msc{gh}} \\
\hspace{1cm}
\dsp + \sum_{M,M',a}\beta_{M,M',a}\ket{L,M}^{SU(2)} \otimes 
   \ket{L,M'}^{SL(2)} \otimes
\chi^a_{-1/2} \ket{0}_f \otimes ce^{-\phi}\ket{0}_{\msc{gh}}~,
\end{array}
\label{phys states}
\end{equation}
where $\ket{L,M}^{SU(2)}$ ($M=L,L-2,\ldots,-L $) and
$\ket{L,M'}^{SL(2)}$ ($M'=L+2, L+4, \ldots$) are the primary
states (zero-mode states)
belonging to $\hat{\cL}_L$ and $\hat{\cD}^+_L$, respectively.
Such physical states are considered in Ref.~\cite{PR}, and it has been claimed
that they correspond to fluctuations around the classical solution 
that make the quadratic variation of the DBI action vanish.

We now concentrate on the 1/2 BPS states, as previously mentioned. 
These are studied in Refs.~\cite{KLL,HS,HHS,AGS} 
as the space-time (anti) chiral 
primaries in the closed string sector.
Among the level-one physical states,
the chiral primaries are obtained by imposing the constraints 
$J^3_0+K^3_0 =0 $ (where $J^A$ and $K^a$ are the total $SL(2;\br)$ and  $SU(2)$
currents including the fermionic contributions), and also the
anti-chiral primaries are given by the constraints $J^3_0-K^3_0 =0$.  
Note that $-J^3_0$ corresponds to the space-time energy operator
(or the space-time conformal weight) and $K^3_0$ evaluates the space-time
R-charge.

The chiral primaries are as follows:\footnote
    {One can explicitly check that these states \eqn{cp 1} and \eqn{acp 1}
     actually preserve a half of space-time SUSY  when using 
     the space-time SUSY generators constructed in Ref.~\cite{GKS}. 
     Moreover, with the help of the Wakimoto free field representation,
     they can be shown to be the (anti) chiral primary states with respect 
     to the {\em full\/} space-time superconformal generators,
     as discussed in Ref.~\cite{HS}.}
\begin{eqnarray}
\ket{L,\psi^{\pm}}^{(+)} &\equiv& \ket{L,\pm L}^{SU(2)} 
 \otimes \ket{L,\mp L \mp 2}^{SL(2)}
 \otimes \psi_{-1/2}^{\pm}\ket{0}_f \otimes c_1e^{-\phi}\ket{0}_{\msc{gh}}
    ~,\hspace{1.5cm}\nn 
\ket{L,\chi^{\pm}}^{(+)}  &\equiv& \ket{L,\pm L}^{SU(2)} \otimes 
 \ket{L,\mp L\mp 2}^{SL(2)}
 \otimes \chi^{\pm}_{-1/2}\ket{0}_f \otimes 
c_1e^{-\phi}\ket{0}_{\msc{gh}} ~.
\label{cp 1}
\end{eqnarray}
Here, $\ket{L,L}^{SU(2)} $ ($\ket{L,- L}^{SU(2)} $) is
the primary state of highest (lowest) weight in the spin $L/2$ 
integrable representation of $\widehat{SU}(2)_{k-2}$, and 
similarly $\ket{L,-L-2}^{SL(2)} $ ($\ket{L,L+2}^{SL(2)}$)
is the primary state of highest (lowest) weight in
$\hat{\cD}^-_L$ ($\hat{\cD}^+_L$) of $\widehat{SL}(2;\br)_{k+2}$.
The anti-chiral primaries are, similarly, given by
\begin{eqnarray}
\ket{L,\psi^{\pm}}^{(-)} &\equiv& \ket{L,\mp  L}^{SU(2)} 
 \otimes \ket{L,\mp L \mp 2}^{SL(2)}
 \otimes \psi_{-1/2}^{\pm}\ket{0}_f \otimes c_1e^{-\phi}\ket{0}_{\msc{gh}} 
   ~,\hspace{1.5cm}\nn 
\ket{L,\chi^{\pm}}^{(-)}  &\equiv& \ket{L,\pm L}^{SU(2)} \otimes 
 \ket{L,\pm L\pm 2}^{SL(2)}
 \otimes \chi^{\pm}_{-1/2}\ket{0}_f \otimes 
c_1e^{-\phi}\ket{0}_{\msc{gh}} ~.
\label{acp 1}
\end{eqnarray}


Next, we consider the sectors with non-trivial winding number
$w$. Generally, the spectral flows do not always 
map an on-shell state to another on-shell state, 
since they do not conserve the BRST charge.
Fortunately, we find that the on-shell (anti) 
chiral states behave nicely under the spectral flow.
To see this, we make use of an idea used in Ref.~\cite{HHS}. 

Recall the spectral flow \eqn{spectral} in the $SL(2;\br)$ sector,
\begin{equation}
\left\{
\begin{array}{l}
 \dsp U_w j^3_n U_w^{-1} = j^3_n + \frac{k+2}{2} w\delta_{n,0} ~,\\ 
 U_w j^{\pm}_n U_w^{-1} = j^{\pm}_{n\mp w} ~.
\end{array}
\right.
\label{spectral SL(2)}
\end{equation} 
We extend the action of the spectral flow operator $U_w$ to the other
sectors as in Ref.~\cite{HHS}:
\begin{equation}
\left\{
\begin{array}{l}
 \dsp U_w^{(+)} k^3_n U_w^{(+)\, -1} 
   = k^3_n -  \frac{k-2}{2} w\delta_{n,0} ~,\\ 
 U_w^{(+)} k^{\pm}_n U_w^{(+)\,-1} = k^{\pm}_{n\mp w} ~,
\end{array}
\right.
\label{spectral SU(2) +}
\end{equation} 
\begin{equation}
\left\{
\begin{array}{l}
 \dsp U_w^{(+)} \psi^3_n U_w^{(+)\,-1} = \psi^3_n ~,\\ 
 U_w^{(+)} \psi^{\pm}_n U_w^{(+)\,-1} = \psi^{\pm}_{n\mp w} ~,
\end{array}
\right.
\label{spectral psi +}
\end{equation} 
\begin{equation}
\left\{
\begin{array}{l}
 \dsp U_w^{(+)} \chi^3_n U_w^{(+)\, -1} = \chi^3_n ~,\\ 
 U_w^{(+)} \chi^{\pm}_n U_w^{(+)\, -1} = \chi^{\pm}_{n\mp w} ~.
\end{array}
\right.
\label{spectral chi +}
\end{equation} 
We also define
\begin{equation}
\left\{
\begin{array}{l}
 \dsp U_w^{(-)} k^3_n U_w^{(-)\, -1} 
   = k^3_n +  \frac{k-2}{2} w\delta_{n,0} ~,\\ 
 U_w^{(-)} k^{\pm}_n U_w^{(-)\,-1} = k^{\pm}_{n\pm w} ~,
\end{array}
\right.
\label{spectral SU(2) -}
\end{equation} 
\begin{equation}
\left\{
\begin{array}{l}
 \dsp U_w^{(-)} \chi^3_n U_w^{(-)\, -1} = \chi^3_n ~,\\ 
 U_w^{(-)} \chi^{\pm}_n U_w^{(-)\, -1} = \chi^{\pm}_{n\pm w} ~,
\end{array}
\right.
\label{spectral chi -}
\end{equation} 
and the action of $U^{(-)}_w$ on the $\psi$ sector is defined to
be the same as that of $U^{(+)}_w$. 

For the $SU(2)$ sector,  \eqn{spectral SU(2) +} and \eqn{spectral SU(2) -} 
simply represent the actions of affine Weyl group, and we find
\begin{equation}
\begin{array}{ll}
 U_w^{(\pm)} ~ :~ \hat{\cL}_L ~ \longrightarrow ~ \hat{\cL}_L ~,
 & (w \in 2\bz)\\
 U_w^{(\pm)} ~:~ \hat{\cL}_L ~ \longrightarrow ~ \hat{\cL}_{k-2-L} ~. 
 & (w \in 2\bz+1) 
\end{array}
\end{equation} 

Now, the important point here is that, as discussed in Ref.~\cite{HHS},
the spectral flow operators $U_w^{(\pm)}$ have the properties  
\begin{eqnarray}
U_w^{(\pm)} Q_{BRST} U_w^{(\pm)\,-1}& =& Q_{BRST} 
-w \Bigg\lb c(0)(J^3\pm K^3)(0) 
\nn && \hspace{1cm}
\left. + \sqrt{\frac{k}{2}}\eta(0)e^{\phi(0)}
 (\psi^3\pm \chi^3)(0)  \right\rb ~,     \\
U_w^{(\pm)} (J^3_0 \pm K^3_0 )U_w^{(\pm)\,-1}
&=& J^3_0 \pm K^3_0~.
\end{eqnarray}
These properties imply that {\em the action of $U^{(+)}_w$ ($U^{(-)}_w$)
is  closed 
in the space of the on-shell space-time (anti) chiral primary states.}
We also note the identities
\begin{equation}
 \begin{array}{l}
 U^{(+)}_{-1}\ket{L,\psi^+}^{(+)} = \ket{k-2-L,\chi^-}^{(+)} ~, ~~~
 U^{(+)}_{-1}\ket{L,\chi^+}^{(+)} = \ket{k-2-L,\psi^-}^{(+)}  ~, \\
 U^{(-)}_{-1}\ket{L,\psi^+}^{(-)} = \ket{k-2-L,\chi^+}^{(-)} ~, ~~~
 U^{(-)}_{-1}\ket{L,\chi^-}^{(-)} = \ket{k-2-L,\psi^-}^{(-)} ~.
\end{array}
\end{equation}  
Using these identities, 
the complete set of on-shell chiral primaries is given by
\begin{eqnarray}
&&\{ ~ \ket{L, w, \psi^+}^{(+)} \equiv U_w^{(+)} \ket{L,\psi^+}^{(+)}, ~
\ket{L, w, \chi^+}^{(+)} \equiv U_w^{(+)} \ket{L,\chi^+}^{(+)} \} ~,
\nonumber\\
&&\hspace{2cm}( w \in \bz ,~ L=0,1,\ldots, k-2)
\label{cp2}
\end{eqnarray}
and for the anti-chiral primaries, we similarly obtain
\begin{eqnarray}
&&\{ ~ \ket{L, w, \psi^+}^{(-)} \equiv U_w^{(-)} \ket{L,\psi^+}^{(-)}~, ~
\ket{L, w, \chi^-}^{(-)} \equiv U_w^{(-)} \ket{L,\chi^-}^{(-)} \} ~.   
\nonumber\\
&&\hspace{2cm}( w \in \bz,~ L=0,1,\ldots, k-2)
\label{acp2}
\end{eqnarray}


To summarize, we have obtained an infinite number of on-shell (anti) chiral
primaries \eqn{cp2} (\eqn{acp2}). The $L$-value undergoes quantum 
truncation that depends on the choice of Cardy states 
of both $SU(2)$ and $SL(2;\br)$ sectors. 
The spectral flows can act on this spectrum
transitively for arbitrary even windings $w$, irrespective of 
the choice of the Cardy states. 
For odd windings, this is generally not the case,
since the spectral flows act as the $\bz_2$-reflection
$L\,\rightarrow\,k-2-L$, on both  
the $\widehat{SU}(2)_{k-2}$  and $\widehat{SL}(2;\br)_{k+2}$ sectors.
Related arguments based on the analysis of classical solutions
are given in Refs.~\cite{PR,LOPT}.

We would like to point out that the spectrum of the 
closed string channel (Cardy states) also includes an infinite number of  
such chiral primary states with arbitrary winding numbers, which can be
constructed in a manner similar to those for the open string spectrum
given by \eqn{cp2} and \eqn{acp2}. 
The essential point is that the  spectral flows of the type
\eqn{spectral outer} preserve the Cardy state of the $SL(2;\br)$ sector 
\eqn{Cardy solution} (up to the signature),
\begin{equation}
 U_w \otimes \tU_w \ket{L}^{SL(2)}_C = (-1)^{Lw} 
\ket{L}^{SL(2)}_C~,
\label{Cardy flow}
\end{equation}
under our construction. 
Similar relations hold also for the other sectors 
(the $SU(2)$ sector and the sectors of free fermions).


\subsection{Comments on boundary states based on continuous series}

The above treatment may be incomplete, because long strings are not
considered  in either the open or closed string channel.
The correct boundary states describing the $AdS_2$-branes 
are expected to have contributions also from the continuous series. 
The components of the boundary states obtained from the discrete series,
which we constructed, describe the open short strings propagating 
in the domain bounded by the $AdS$ radius, and all the space-time 
chiral primary states (as both open and closed strings)
appear in this sector.

It is not difficult to construct the Ishibashi states for 
the continuous series.  However, there is a large ambiguity when 
we attempt to construct the Cardy states from them.  We could 
expect open string excitations of the following types to result 
from such Cardy states constructed from the continuous series:
\begin{description}
 \item[(1)] open long strings with arbitrary winding number $w$,
 \item[(2)] open short strings with arbitrary winding number $w$
  that propagate in the domain $\rho \gsim 1$ (the $AdS$ length).
\end{description}

The classical solutions of open long strings with arbitrary winding
numbers are explicitly constructed in Ref.~\cite{LOPT}, and it is natural
to expect the corresponding excitations (1) in the quantum open string 
spectrum.   It can be easily shown from consideration of the modular weight 
that we must use boundary states constructed from the continuous series 
with {\em no winding\/} in order to obtain the summation over windings 
in the open string channel.   Conversely, if we start with 
the boundary states with the summation over windings,
we obtain the open string spectrum with no winding. 
Therefore, it seems difficult to naively construct  the Cardy states
so as to be consistent with the spectral flow symmetry
in {\em both\/} open and closed string channels in this case.
This is one of the main puzzles of this study and we need further detailed 
investigations to reach a definite conclusion.

To assume the second excitations (2) is also quite natural, 
since the Cardy states considered here are expected to interact with 
the strings that can propagate in the region far from the center, 
in contrast to those we discussed in the previous subsections.
In fact, choosing some hyperbolic functions as the wave functions, 
we can obtain the open string spectrum of the discrete series, as presented 
in Ref.~\cite{GKSch}. However, at least with a naive consideration,
we would face the following difficulties if we start with 
the criterion used in the previous arguments:
\begin{enumerate}
 \item It seems difficult to incorporate the non-trivial winding sectors
       in the open string channel.
 \item It seems difficult to make the open string spectrum compatible
       with the unitarity bound. 
\end{enumerate}  

For the reasons given above, it is unclear 
at this point whether we can reproduce a consistent open string spectrum 
from the boundary states constructed with the continuous series. 
This is an important problem, which should be resolved 
in future studies.\footnote
  {Recently, this problem has been discussed in Ref.~\cite{RR}.}


\section{Discussion}

In this paper we have studied the $AdS_2$-branes in the string 
theory on the $AdS_3$ background from the viewpoint of boundary states,
emphasizing the role of open-closed duality in string theory.
We have constructed the Cardy states from the discrete series.
These states possess  the following desirable properties:
\begin{enumerate}
\item They are compatible with the symmetry of the spectral flow. 
\item They are consistent with the unitarity and normalizability
      condition in the open string spectrum.
\end{enumerate}

We have found that the first property above yields a rich structure of
physical BPS states both in the open and closed string channels;
that is, the spectral flows consistently act on the spectrum
of infinitely many space-time chiral primaries. 
Such physical BPS states are believed to play important roles 
in the context of $AdS_3/CFT_2$ correspondence. One interesting 
direction for future works is the attempt to describe
the D-branes in $AdS_3$ string theory in the framework of
the boundary CFT (in the sense of $AdS/CFT$ correspondence 
though this terminology is sometimes very confusing) rather 
than the world sheet CFT approach.  The analysis of physical BPS
states given in this paper should provide helpful insights for such studies.

The second property above originates from the quantum truncation, like
the fusion rule in the $SU(2)$ WZW model. The existence of this truncation
seems to lead to the ``fuzziness'' of brane dynamics, as in the 
$SU(2)$ case. Such fuzziness may seem peculiar, since $AdS_3$ 
and $AdS_2$ are non-compact spaces, in contrast to $SU(2)$ and $S^2$. 
In fact, the classical analysis given in Ref.~\cite{LOPT} suggests 
that we should not have such a truncation in the open string spectrum. 
One possibility to resolve this apparent contradiction may be to claim that 
our Cardy states \eqn{Cardy solution}
{\em define\/}  ``fuzzy $AdS_2$-branes'' that {\em do not have
counterparts in the classical brane geometry.\/}
More modestly speaking,  \eqn{Cardy solution}
should correspond to the component of the Cardy states 
obtained from the discrete series, and
we will have to further take account of the component obtained from the 
continuous series  
in order to construct the complete Cardy states describing 
the ``classical'' $AdS_2$-branes.

As discussed, the Cardy states  \eqn{Cardy solution} 
can interact only with the short strings propagating in the domain
bounded by the $AdS$ radius. 
This fact is supposed to be the origin of the fuzziness mentioned above.
Also it is natural to hypothesize that 
the component obtained from the continuous series effectively describes 
(1) open long strings that can reach the asymptotic region (the
boundary of $AdS_3$), and 
(2) open short strings propagating in region far from the center.
If this is indeed the case, we will be able to obtain boundary states
reproducing the classical geometry of $AdS_2$-branes. 
However, as we mentioned in the last part of section 3, there are several
puzzles concerning the Cardy states constructed from the continuous series,
and they have to be resolved in future studies.


\section*{Acknowledgements}

We would like to thank K. Hosomichi and Y. Satoh for 
helpful conversations.

The work of Y. S. is supported in part by a
Grant-in-Aid for the Encouragement of Young Scientists 
($\sharp 13740144$) 
and also by a Grant-in-Aid for Scientific Research on Priority Area 
($\sharp 707$), 
``Supersymmetry and Unified Theory of Elementary Particles,'' 
both from the Japanese Ministry of Education, Culture, Sports, Science and 
Technology.



\appendix
\section{Notation}

The theta functions are defined as follows
(where we define $q\equiv e^{2\pi i \tau}$ and  $y\equiv e^{2\pi i z}$):
\begin{equation}
\begin{array}{l}
 \dsp \th_1(\tau,z) =i\sum_{n=-\infty}^{\infty}(-1)^n q^{(n-1/2)^2/2} y^{n-1/2}
 \\ \hspace{1.3cm}
 \equiv 2 \sin(\pi z)q^{1/8}\prod_{m=1}^{\infty}
    (1-q^m)(1-yq^m)(1-y^{-1}q^m), \\
 \dsp \th_2(\tau,z)=\sum_{n=-\infty}^{\infty} q^{(n-1/2)^2/2} y^{n-1/2}
 \\ \hspace{1.3cm}
 \equiv 2 \cos(\pi z)q^{1/8}\prod_{m=1}^{\infty}
    (1-q^m)(1+yq^m)(1+y^{-1}q^m), \\
 \dsp \th_3(\tau,z)=\sum_{n=-\infty}^{\infty} q^{n^2/2} y^{n}
  \equiv \prod_{m=1}^{\infty}
    (1-q^m)(1+yq^{m-1/2})(1+y^{-1}q^{m-1/2}), \\
 \dsp \th_4(\tau,z)=\sum_{n=-\infty}^{\infty}(-1)^n q^{n^2/2} y^{n}
  \equiv \prod_{m=1}^{\infty}
    (1-q^m)(1-yq^{m-1/2})(1-y^{-1}q^{m-1/2}) ~,
 \end{array}
 \end{equation}
and also, 
\begin{equation}
 \Th{m}{k}(\tau,z)=\sum_{n=-\infty}^{\infty}
 q^{k(n+\frac{m}{2k})^2}y^{k(n+\frac{m}{2k})} .
\end{equation}
We also set
\begin{equation}
\eta(\tau)=q^{1/24}\prod_{n=1}^{\infty}(1-q^n)~,
\end{equation}
and often use the identity
\begin{equation}
 \partial_z \th_1(\tau, z)|_{z=0} = 2\pi \eta(\tau)^3~.
\end{equation}


\section{Some Mathematical Comments}

The aim of this appendix is to remove the mathematical subtlety 
in our argument in section 3. 
Here, it is useful to introduce the definition
\begin{equation}
 f_a(x) \df 
   \left(\frac{a}{\pi}\right)^{\frac{1}{4}}\, e^{- \frac{1}{2}a x^2}~,
\end{equation}
whose Fourier transform is given by $f_{1/a}$.
For positive real $a$, the Poisson resummation formula leads to
\begin{equation}
\sum_{n\in\bsz}\,f_a(nL)= 
\frac{\sqrt{2 \pi}}{L}\sum_{m\in\bsz}\,f_{1/a}(2 \pi m/L)~, ~~~(L>0)
\label{poisson}
\end{equation}
which gives us the modular transformation formula
of theta functions. However, since we are employing the world sheet with 
the Lorentzian signature in this paper, we face the case in which 
$a \in i \br$.
In this case, the formula \eqn{poisson} is not correct in a naive sense, 
because the power series on both sides do not absolutely converge.
Nevertheless, we interpret it correctly meaning as an identity
among the  {\em generalized functions} (or {\em
distributions}) \cite{Gelfand}.  Now, we briefly sketch how it works.

Let us consider the space of ``rapidly decreasing functions''
$\cS(\br)$. 
More precisely, $f\in \cS(\br)$ implies that $f$ is a ($\bc$-valued)
smooth function such that 
\begin{eqnarray}
p_{m,n}(f)  &<& +\infty ~, ~~(\forall m, n \in \bz_{\geq 0}) \nonumber \\
p_{m,n}(f)  & \df & \max_{\stackrel{\al\leq m}{\beta \leq n}}
\, \sup_{x\in\bsr} \, |x^{\al}\partial^{\beta}f(x)|~.
\end{eqnarray}
The function space $\cS(\br)$ is a countably normed space with respect to
$\{p_{m,n}\}_{m,n\in \bsz_{\geq 0}}$,
and  the generalized functions should be defined as the elements of
$\cS(\br)^*$, i.e. continuous linear functionals on $\cS(\br)$.
The Fourier transform $\tilde{F}$ of $F\in \cS(\br)^*$ is 
defined by the relation
\begin{equation}
\langle \tilde{F}, \tilde{f}\rangle = \langle F, f \rangle ~,
\end{equation}
for arbitrary $f \in \cS(\br)$, and $\tilde{f}$ is its Fourier 
transform. 
In this sense, $f_{ia}$ ($a\in \br$) can be regarded as a generalized 
function, and its Fourier transform is equal to $f_{1/ia}$. 
The only non-trivial fact we use to prove this
is that we have a dense subset of $\cS(\br)$ composed of the functions 
of the form $P(x)e^{-\frac{1}{2}x^2}$, where $P(x)$ represents 
arbitrary polynomial (Hermite functions, essentially).

We can further obtain the identity 
\begin{equation}
\sum_{n\in\bsz}\,f_{ia}(x+nL)= 
\frac{\sqrt{2 \pi}}{L}\sum_{m\in\bsz}\,f_{1/ia}(2 \pi m/L)
  e^{2\pi i \frac{mx}{L}}~.
\label{poisson 2}
\end{equation}
Both sides of this equation are well-defined as the generalized
functions associated with 
the ``periodic function version'' of $\cS(\br)$. To show this, we define
we set $\cS_L$ with period $L$ as
\begin{equation}
\cS_L \df \left\{f(x)= \sum_{n\in\bsz}\,a_n e^{2\pi i \frac{nx}{L}}~;~
 \lim_{|n|\rightarrow +\infty}\, |n^ma_n|=0~,~ 
  (\forall m \in \bz_{\geq 0}) \right\}~,
\end{equation}
and both sides of \eqn{poisson 2} are then seen to be well-defined as
elements of $\cS_L^*$, owing  to  the property 
$ \left|\int_{-L/2}^{L/2}dx\, F(x)\bar{f}(x)\right| < +\infty$ for
arbitrary $f\in \cS_L$. 
In particular, take an arbitrary  series 
$\{\rho_{\nu}\}_{\nu \in \bsz_{> 0}} \subset \cS_L$
such that $ \lim_{\nu \rightarrow +\infty}\, \rho_{\nu} (x)
= \sum_{n\in\bsz}\,\delta(x+ n L)$ (for example, 
$ \rho_{\nu} (x):= \sum_{n\in \bsz} \, \sqrt{\frac{\nu}{\pi}}\, 
e^{-\nu(x+nL)^2} $), then we obtain 
\begin{equation}
\sum_{n\in\bsz}\,\int_{-L/2}^{L/2}\,dx\, \rho_{\nu}(x) f_{ia}(x+nL)= 
\frac{1}{L}\sum_{m\in\bsz}\,\int_{-L/2}^{L/2}\,dx\, \rho_{\nu}(x) 
f_{1/ia}(m/L)e^{2\pi i \frac{mx}{L}}~,
\label{poisson 3}
\end{equation}
for arbitrary $\nu \in \bz_{>0}$. This is a natural generalization of
\eqn{poisson} 
and precisely the identity we want. 
The identities of the modular transformations presented in section 3 
should be understood in this manner.

\vspace{2cm}



\end{document}